\documentstyle[twocolumn,prb,aps,floats,psfig]{revtex} 
\begin{document}
\twocolumn[\hsize\textwidth\columnwidth\hsize\csname 
@twocolumnfalse\endcsname
\title{Nonadiabatic Pauli susceptibility in fullerene compounds} 
\author{E. Cappelluti$^1$, C. Grimaldi$^2$ and L. Pietronero$^1$} 
\address{$^1$Dipartimento di Fisica, Universit\'{a} di Roma I ``La
Sapienza", 
Piazzale A.  Moro, 2, 00185 Roma, Italy \\
Istituto Nazionale Fisica della Materia, Unit\'a di Roma 1, Italy}
\address{$^2$ D\'epartement de Microtechnique - IPM,
\'Ecole Polytechnique F\'ed\'erale de Lausanne, CH-1015 Lausanne, 
Switzerland}
\date{\today} 
\maketitle 
\begin{abstract}
Pauli paramagnetic susceptibility $\chi$ is unaffected by
the electron-phonon interaction in the Migdal-Eliashberg context.
Fullerene compounds however do not fulfill the adiabatic assumption
of Migdal's theorem and nonadiabatic effects are expected to be relevant
in these materials. In this paper we investigate the Pauli
spin susceptibility in nonadiabatic regime
by following a conserving approach based on Ward's
identity. We find that a sizable renormalization of $\chi$ due
to electron-phonon coupling appears when nonadiabatic effects are taken
into account.
The intrinsic dependence of $\chi$ on the electron-phonon interaction
gives rise to a finite and negative isotope effect which could
be experimentally detected in fullerides.
In addition, we find an enhancement of the spin susceptibility with
temperature increasing, in agreement with the 
temperature dependence of $\chi$ observed in fullerene compounds.
The role of electronic correlation is also discussed.
\\

PACS number(s): 74.70.Wz, 71.38.-k, 63.20.Kr

\end{abstract}

\vskip 2pc 
] 
\narrowtext

\section{Introduction}
\label{intro}

The Pauli susceptibility, $\chi$, of wide-band metals is usually
considered as unrenormalized by electron-phonon effects.
As pointed out some time ago by Fay and Appel,\cite{fay} $\chi$ is
actually affected by the electron-phonon coupling, the effect
having magnitude of order $\lambda\omega_{\rm ph}/E_F$, where
$\lambda$ is the electron-phonon coupling, $\omega_{\rm ph}$ is a typical
phonon frequency and $E_F$ is the Fermi energy. For wide-band metals,
$\omega_{\rm ph}\ll E_F$ and the electron-phonon renormalization of
$\chi$ is therefore negligible.
This result basically follows from the observation that the lowest
order electron-phonon correction to $\chi$ is a
vertex diagram which,  according to Migdal's theorem,\cite{migdal} is of order
$\lambda\omega_{\rm ph}/E_F$. Hence the absence of phonon
effects in $\chi$ is just a manifestation of the electron-phonon
adiabaticity of normal metals. 

This situation may be somewhat different for weak ferromagnetic
metals like ZrZn$_2$.\cite{fay,enz} In this case, 
the electron-phonon corrections
compete with the exchange term leading to a possible phonon-dependent
ferromagnetic transition temperature. Experiments have however reported
only small isotope effects in ZrZn$_2$ with quite large error 
bars,\cite{knapp} 
leaving the problem of phonon-corrected weak-ferromagnetism  
essentially unresolved. The recent discovery of giant 
isotope shifts in manganites proves however the existence of 
ferromagnetic materials with important electron-phonon effects 
in their magnetic properties.\cite{zhao} 

The possibility of having electron-phonon dependent spin susceptibility is 
however not correlated exclusively to the vicinity of magnetic instabilities. 
In fact
as long as  $\lambda\omega_{\rm ph}/E_F$ is not negligible, as in very
narrow band systems, the electron-phonon vertex contributions are
no longer unimportant so that $\chi$ is expected to acquire a 
phonon renormalization. The most promising candidates for the
observation of this effect are the C$_{60}$-based materials. Fullerene
compounds have in fact phonon modes extending up to $0.2$ eV and
Fermi energies of about $0.3$ eV.\cite{gunny} 
Therefore, $\omega_{\rm ph}/E_F$
is in principle large suggesting that C$_{60}$-based metals could be labeled
as nonadiabatic systems.

Recently, the finding of a superconducting transition at
$T_c=52$ K in hole doped C$_{60}$ single crystals has rised a
renewed interest in these materials.\cite{batlogg} Nonzero isotope effects
and other properties strongly indicates that superconductivity
in fullerides is driven by electron-phonon interaction.\cite{gunny}
However, the description of superconductivity of Rb$_3$C$_{60}$
within the traditional Migdal-Eliashberg (ME) theory is found to be
inconsistent with respect to the adiabatic hypothesis 
$\lambda\omega_{\rm ph}/E_F\ll 1$
which is at the basis of the ME theory itself.\cite{cgps}
Instead, by relaxing the adiabatic hypothesis, a generalized formulation
which includes nonadiabatic channels in the electron-phonon
interaction provides a more self-consistent decription of superconductivity,
and suggests that the key ingredient for the high values
of $T_c$ in C$_{60}$-based materials is a constructive nonadiabatic 
interference rather than strong electron-phonon 
couplings.\cite{cgps,pietro1}

In principle, the hypothesis that superconductivity in fullerene compounds
is enhanced essentially by nonadiabatic electron-phonon effects
can be sustained by the observation of independent signatures of 
nonadiabaticity. In this respect and according to what we have pointed
out at the beginning,  Pauli susceptibility is a
quantity where such signatures could be found.
In this paper, we provide extensive calculations of $\chi$
beyond the adiabatic limit by including nonadiabatic effects at different
stages of a perturbation theory in $\lambda\omega_{\rm ph}/E_F$.
We find that when $\omega_{\rm ph}/E_F$ is no longer negligible, (i)
the paramagnetic spin susceptibility can be considerably reduced
with respect to the adiabatic limit, (ii) it acquires
a negative isotope effect and (iii) a possible anomalous temperature 
dependence at constant sample volume. These features are signatures
of nonadiabatic electron-phonon interaction and prediction (ii) and (iii)
are susceptible of experimental verification.

\section{Pauli susceptibility by Ward's identity}
\label{paulisus}

As remarked in the introduction, the effect of electron-phonon
interaction on spin susceptibility has already attracted some interest
in the past mainly in relation to weak antiferromagnets. However,
different approaches led
to different results,\cite{fay,enz}
reflecting the lacking of a controlled theory.

In our paper we use the functional formalism based on the Baym-Kadanoff
technique to derive a conserving derivation of Pauli susceptibility
valid for both electron-phonon and electron-electron interactions. 
The Pauli susceptibility $\chi$
is calculated by the knowledge of the spin vertex function which
is related to the self-energy via a Ward's identity.

Following the Baym-Kadanoff technique,
we introduce an external magnetic field $H$ coupled with the electrons
which induces a magnetization $M$.
The interaction hamiltonian describing the coupling
of $H$ with the electron spins is:
\begin{equation}
H_h=-h\sum_{{\bf k},\sigma}\sigma c^{\dagger}_{{\bf k}\sigma}
c_{{\bf k}\sigma},
\label{ham1}
\end{equation}
where $h=\mu_{\rm B} H$ and 
$c^{\dagger}_{{\bf k}\sigma}$ ($c_{{\bf k}\sigma}$) are
creation (annihilation) fermionic operators for electrons with
wave number ${\bf k}$ and spin index $\sigma=\pm 1$. 
The electron magnetization $M$ due to (\ref{ham1}) is given by 
$M=\mu_{\rm B}\sum_{\sigma}\sigma n_{\sigma}$ where 
$n_{\sigma}=\sum_{{\bf k}}\langle c^{\dagger}_{{\bf k}\sigma}
c_{{\bf k}\sigma}\rangle$ and $\langle\cdots\rangle$ denotes the
statistical average. We express now the magnetization $M$ in terms
of the finite temperature single electron propagator
\begin{equation}
G_{\sigma}({\bf k},\tau)=-\langle T_{\tau}
c_{{\bf k}\sigma}(\tau)c^{\dagger}_{{\bf k}\sigma}(0)\rangle ,
\label{green1}
\end{equation}
where $T_{\tau}$ is the time ordering operator and $\tau$ is the
imaginary time. Since $\langle c^{\dagger}_{{\bf k}\sigma}
c_{{\bf k}\sigma}\rangle=G_{\sigma}({\bf k},0^{-})$, the magnetization
can be expressed as:
\begin{eqnarray}
M&=&\mu_{\rm B}\sum_{{\bf k},\sigma}\sigma 
G_{\sigma}({\bf k},0^{-}) \nonumber \\
&=&\mu_{\rm B}T\sum_n\sum_{{\bf k},\sigma}\sigma
G_{\sigma}({\bf k},n)e^{-i\omega_n 0^{-}} .
\label{emme1}
\end{eqnarray}
In the above expression, $G_{\sigma}({\bf k},n)$ is the 
thermal electron propagator which satisfies the following Dyson equation:
\begin{equation}
G_{\sigma}^{-1}({\bf k},n)=i\omega_n-\epsilon({\bf k})+\mu+h\sigma-
\Sigma_{\sigma}({\bf k},n),
\label{greendys}
\end{equation}
where $\omega_n=(2n+1)\pi T$, $n=0,\pm 1,\pm 2,\ldots$, are Matsubara frequencies,
$\epsilon({\bf k})$ the electron dispersion, $\mu$
the chemical potential and $\Sigma_{\sigma}({\bf k},n)$ is
the electronic self-energy due to the coupling to phonons and
to the electron-electron interaction.
The spin susceptibility $\chi$ is formally
given by $M=\chi H$ where $\chi=[dM/dH]_0$ is the derivative
of the magnetization at zero field. Hence, from Eq. (\ref{emme1}),
a general expression for $\chi$ is the following:
\begin{eqnarray}
\label{chi1}
\chi(T)&=&\mu_{\rm B}^2T\sum_n\sum_{{\bf k},\sigma}\sigma
\left[\frac{d G_{\sigma}({\bf k},n)}{dh}\right]_0 \nonumber \\
&=&-2\mu_{\rm B}^2T\sum_n\sum_{{\bf k}}
G({\bf k},n)^2\Gamma({\bf k},n),
\label{eqchi}
\end{eqnarray}
where $G({\bf k},n)$ is the electron propagator for
zero magnetic field which satisfies the $H\rightarrow 0$ limit
of Eq. (\ref{greendys}):
\begin{equation}
G^{-1}({\bf k},n)=i\omega_n-\epsilon({\bf k})+\mu-
\Sigma({\bf k},n).
\label{green1bis}
\end{equation}
In the second term of Eq. (\ref{chi1}) we have introduced the
spin-vertex function:
\begin{equation}
\Gamma({\bf k},n)=\frac{1}{2}\sum_{\sigma}\sigma
\left[\frac{d G_{\sigma}^{-1}({\bf k},n)}{dh}\right]_0 
\label{gamma1}
\end{equation}
In equations (\ref{eqchi}) and (\ref{gamma1}), the notation
$[\cdots]_0$ indicates that the quantity in brackets must be
calculated for zero magnetic field.
Plugging Eq. (\ref{greendys})
into (\ref{gamma1}) the resulting vertex $\Gamma$ satisfies
the Ward's identity:
\begin{equation}
\Gamma({\bf k},n)=1-\frac{1}{2}\sum_{\sigma}\sigma
\left[\frac{d \Sigma_{\sigma}({\bf k},n)}{dh}\right]_0
\label{gamma2}
\label{ward}
\end{equation}
and, since $\Sigma_{\sigma}({\bf k},n)$ is a functional
of the electron propagator, a self-consistent relation between
(\ref{gamma1}) and (\ref{gamma2}) is obtained which permits 
to calculate the spin susceptibility. 
At this point, the spin susceptibility $\chi$ can be calculated
once the electron self-energy and its magnetic field
dependence is known.

\section{Nonadiabatic Pauli susceptibility}

Electron-phonon interaction is usually neglected in the calculations
of the Pauli susceptibility although it can strongly renormalize other
physical quantities.\cite{grimvall}
Indeed it can be shown that the electron-phonon
self-energy depends on the external magnetic field $h$ at least as
\begin{equation}
\label{lim}
\lim_{h \rightarrow 0}
\Sigma^{\rm ep}({\bf k},n) \sim h \, O(\omega_{\rm ph}/E_{\rm F}),
\end{equation}
where $\omega_{\rm ph}$ characterizes the phonon frequency scale
and $E_{\rm F}$ is the Fermi energy.
In wide-band materials the adiabatic ratio
$\omega_{\rm ph}/E_{\rm F} \ll 1$
and the electron-phonon contribution to $\chi$ can be
consequently disregarded. From Eq. (\ref{lim}), we see
that electron-phonon effects appear in $\chi$
only at a nonadiabatic level.
Hence, any evidence of electron-phonon
effects on the Pauli susceptibility is thus a direct proof
of a nonadiabatic electron-phonon coupling.

In order to properly include electron-phonon interactions in $\chi$
we need then to explicitely specify the functional form
of the nonadiabatic electron-phonon self-energy.
Nonadiabatic effects enter in a twofold way in the electron-phonon
self-energy: finite bandwidth effects and vertex diagrams.
These two kinds of effects, of course, are of the same order
and there is no justification for neglecting vertex corrections
with respect to finite bandwidth effects.

In this paper we consider two approximation schemes for the electron-phonon
self-energy. The first one is essentially the mean-field theory which
corresponds to the non-crossing approximation. 
It is diagrammatically equivalent to the 
Migdal-Eliashberg electron-phonon self-energy without however assuming
an infinite electron bandwidth compared to the relevant phonon energies.
At this level only
finite bandwidth nonadiabatic effects are considered.
The second one includes first order electron-phonon
vertex corrections as well as
finite bandwidth effects in the framework on the nonadiabatic Fermi liquid
picture.\cite{pietro1}
Both approximation schemes reduce to the adiabatic Migdal-Eliashberg
limit for $\omega_{\rm ph}/E_F\rightarrow 0$.

\subsection{Non-crossing self-energy}
\label{non-cross}

\begin{figure}[t]
\centerline{\psfig{figure=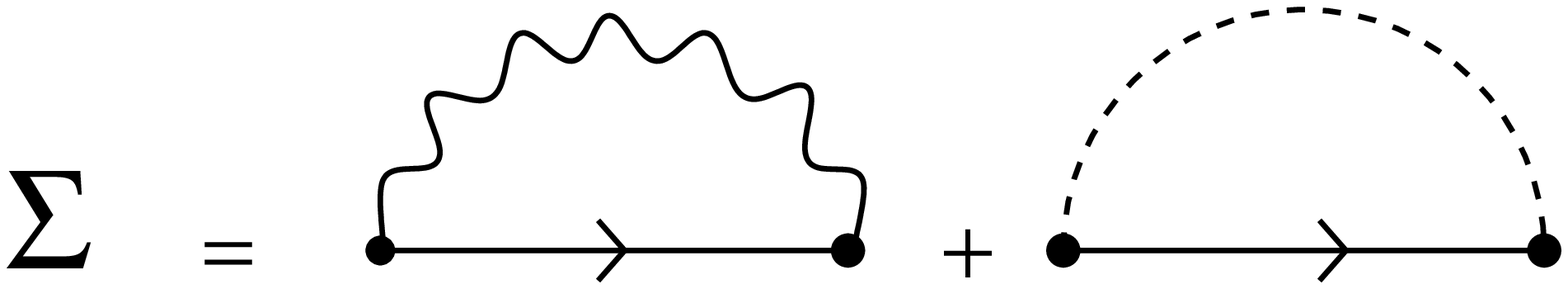,width=6.5cm}}
\caption{Electronic self-energy within the non-crossing approximation.}
\label{figselfen-nca}
\end{figure}

The self-consistent non-crossing approximation neglects
the vertex corrections in the electron-phonon self-energy and
for $\omega_{\rm ph}/E_F\ll 1$ reduces to the ME theory of 
the electron-phonon coupled system. 
The diagrammatic representation
of the electron self-energy is shown in Fig. \ref{figselfen-nca}
where the wiggled line
represents the phonon propagator 
and the dashed line the electron-electron Coulomb repulsion.
The corresponding compact expression of the non-crossing
self-energy reads:
\begin{equation}
\Sigma_{\sigma}(k)=\sum_{ k'}\left[
V(k-k')+I e^{-i\omega_m 0^-}\right]G_{\sigma}(k'),
\label{self1}
\end{equation}
where $k$ and $k'$  are fermionic four-vectors defined as
$k\equiv ({\bf k},i\omega_n)$ and $k'\equiv ({\bf k}',i\omega_m)$. 
Moreover $\sum_k\equiv -T\sum_n\sum_{{\bf k}}$ and
$V(k-k')\equiv |g({\bf k}-{\bf k}')|^2D(k-k')$, where
$g({\bf k}-{\bf k}')$ is the electron-phonon matrix element and
$D(k-k')$ is the phonon propagator. $I$ is exchange Coulomb interaction
which gives rise to the Stoner enhancement factor.

\begin{figure}
\centerline{\psfig{figure=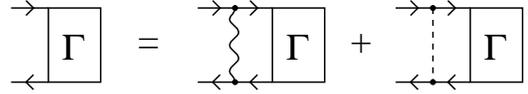,width=7.5cm}}
\caption{Diagramatic self-consistent expression of the
spin vertex in non-crossing approximation.}
\label{figvertexchi-nca}
\end{figure}

By introducing Eq. (\ref{self1}) into the expression of the
spin-vertex function (\ref{gamma2}) and using Eq. (\ref{gamma1}) we obtain:
\begin{eqnarray}
\label{gamma3}
\Gamma(k)&=&1-\sum_{k'}\left[V(k-k')+I e^{-i\omega_m 0^-}\right]
\sum_{\sigma}\frac{\sigma}{2}\left[\frac{dG_{\sigma}(k')}{dh}\right]_0
\nonumber \\
&=&1+\sum_{k'}[V(k-k')+I]G(k')^2\Gamma(k').
\label{eqga}
\end{eqnarray}
The non-crossing approximation for the self-energy leads therefore
to a self-consistent ladder equation for $\Gamma({\bf k},n)$\cite{grima2} 
(Fig. \ref{figvertexchi-nca}). 
If we consider dispersionless phonons of frequency $\omega_0$, the
above ladder equations can be rewritten in the extended notation as:
\begin{eqnarray}
\label{gamma4}
\Gamma({\bf k},n)&=&1-T\sum_{{\bf k}',m}
\left[|g({\bf k}-{\bf k}')|^2D(n-m)+I\right] \nonumber \\
&&\times G({\bf k}',m)^2 \Gamma({\bf k}',m),
\end{eqnarray}
where
\begin{equation}
\label{phonon1}
D(n-m)=\frac{\omega_0^2}{(i\omega_m-i\omega_m)^2-\omega_0^2}.
\end{equation}

The electron propagator appearing in Eq. (\ref{eqga})
satisfies the Dyson equation (\ref{green1bis}) with the zero field limit
of the self-energy (\ref{self1}):
\begin{eqnarray}
\Sigma({\bf k},n)&=&-T\sum_{{\bf k}',m} 
|g({\bf k}-{\bf k}')|^2D(n-m) G({\bf k}',m)\nonumber\\
&&+I T\sum_{{\bf k}',m} G({\bf k}',m) e^{-i\omega_m 0^-},
\label{self2}
\end{eqnarray}
In the simplest case of ${\bf k}$-independent Coulomb repulsion
here considered the second term on the right side of
Eq. (\ref{self2}) gives rise just to a constant term
which can be absorbed into a redefinition of the chemical potential
$\mu' \rightarrow \mu$.
We can then neglect it since we shall consider only half-filled
systems for which we set $\mu'=0$.

In an isotropic system the angular dependence of the self-energy
and of the vertex function in equations (\ref{gamma4}) and (\ref{self2})
is negligible and it can be dropped.
Therefore, following
the same procedure reported in Ref. \onlinecite{grima2},
we replace the electron-phonon interaction $|g({\bf k}-{\bf k}')|^2$
by its average over the Fermi surface:
\begin{equation}
\label{ave1}
|g({\bf k}-{\bf k}')|^2\rightarrow \langle\langle
|g({\bf k}-{\bf k}')|^2\rangle\rangle_{\rm FS}\equiv g^2,
\end{equation}
where
\begin{equation}
\label{ave2}
\langle\langle|g({\bf k}-{\bf k}')|^2\rangle\rangle_{\rm FS}
=\frac{\sum_{{\bf k},{\bf k}'}|g({\bf k}-{\bf k}')|^2
\delta[\epsilon({\bf k})]\delta[\epsilon({\bf k}')]}
{\sum_{{\bf k},{\bf k}'}\delta[\epsilon({\bf k})]\delta[\epsilon({\bf k}')]}.
\end{equation}
In this way, the electron self-energy becomes independent of the
momentum, $\Sigma({\bf k},n)=\Sigma(n)$, 
and the electron propagator becomes at half-filling:
\begin{equation}
\label{green2}
G({\bf k},n)=\frac{1}{iW_n-\epsilon({\bf k})},
\end{equation}
where we have set $\Sigma(n)=i\omega_n-iW_n$. By using a constant density
of states, $N_0$, the momentum summation in Eq. (\ref{self2}) is transformed
as $\sum_{{\bf k}'}\rightarrow N_0\int^{+E_F}_{-E_F}d\epsilon$ and, under
integration over $\epsilon$, the renormalized frequency $W_n$ becomes:
\begin{equation}
\label{self3}
W_n=\omega_n-\lambda\pi T\sum_mD(n-m)
\frac{2}{\pi}\arctan\left(\frac{E_F}{W_m}\right),
\end{equation}
where $\lambda=N_0 g^2$ is the electron-phonon coupling constant.
By using (\ref{ave1}) and (\ref{green2}), also the spin-vertex function
(\ref{gamma4}) becomes momentum independent and, by following the
same steps as above, it reduces to:
\begin{equation}
\label{gamma6}
\Gamma(n)=1+T\sum_m
\left[\lambda D(n-m)+I\right]\frac{2E_F}{W_m^2+E_F^2}\Gamma(m).
\label{gammanca}
\end{equation}
Finally, since the self-energy and the spin-vertex function depend
only on the frequency, the summation over ${\bf k}$ can be
readily performed in (\ref{chi1}) leading to:
\begin{equation}
\label{chi5}
\chi(T)=\chi_{\rm P} T\sum_n\frac{2E_F}{W_n^2+E_F^2}
\Gamma(n),
\end{equation}
where $\chi_{\rm p}=2\mu_{\rm B}^2N_0$. The spin susceptibility
is then obtained by the solution of equations (\ref{self3}), (\ref{gamma6}),
and (\ref{chi5}).

\subsection{Vertex corrected self-energy}
\label{vercor}

\begin{figure}
\centerline{\psfig{figure=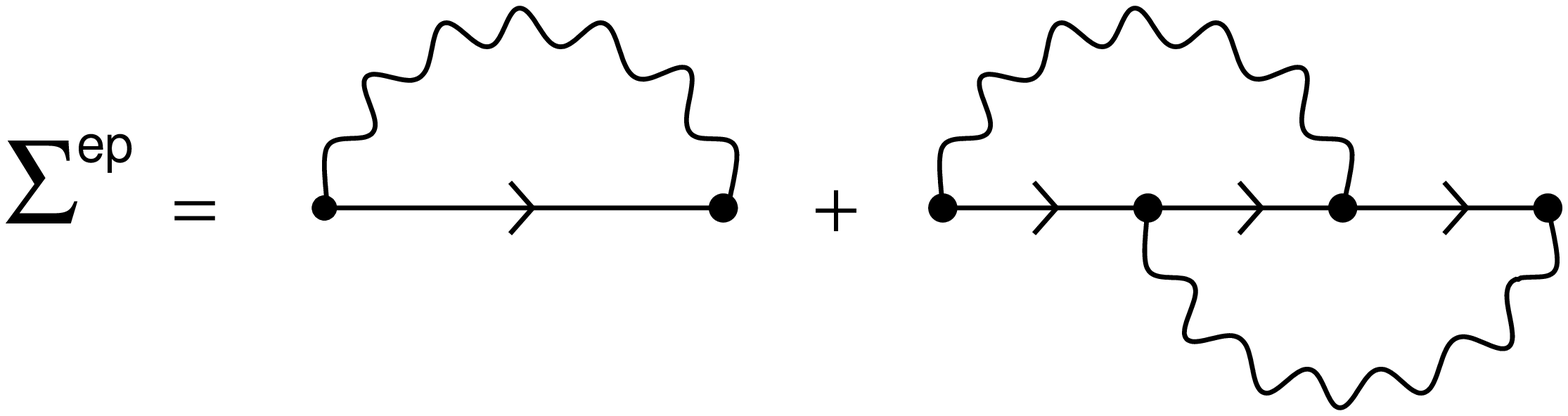,width=7.2cm}}
\caption{Vertex corrected electron-phonon self-energy.}
\label{figselfen-nafl}
\end{figure}

\begin{figure}
\centerline{\psfig{figure=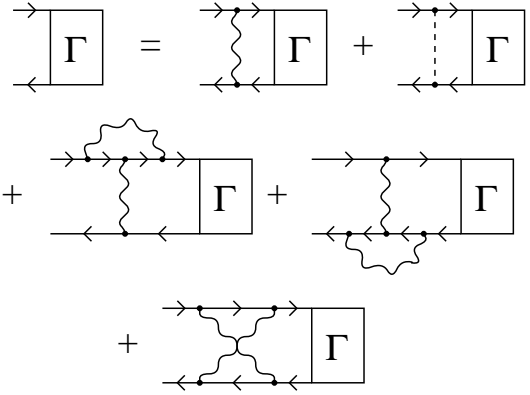,width=7.5cm}}
\caption{Diagramatic self-consistent expression of the
spin vertex in the vertex corrected theory.}
\label{figvertexchi-nafl}
\end{figure}

In the vertex correction approximation, the electron-phonon self-energy
$\Sigma^{\rm ep}$ is modified with respect to the ME one by the
inclusion of the first electron-phonon vertex diagram as
shown in Fig. \ref{figselfen-nafl}.
By making use of the condensed notation introduced in Sec. \ref{non-cross},
the vertex corrected self-energy can therefore be written as follows:
\begin{eqnarray}
\label{self4}
\Sigma^{\rm ep}_{\sigma}(k)&=&\sum_{k'}V(k-k')G_{\sigma}(k') \nonumber \\
&\times &\left[1+\sum_{q}V(k-q)G_{\sigma}(q-k+k')G_{\sigma}(q)\right].
\end{eqnarray}
where $q\equiv ({\bf q},i\omega_l)$.
The derivative of $\Sigma^{\rm ep}_{\sigma}(k)$ with respect to
$h=\mu_{\rm B}H$ calculated at zero magnetic field is:
\begin{eqnarray}
\label{self5}
&&\left[\frac{d\Sigma^{\rm ep}_{\sigma}(k)}{dh} \right]_0=
\sum_{k'}V(k-k')\left[\frac{dG_{\sigma}(k')}{dh}\right]_0 \nonumber \\
&&+\sum_{k',q}V(k-k')V(k-q)G(q-k+k')G(q)
\left[\frac{dG_{\sigma}(k')}{dh}\right]_0 \nonumber \\
&&+\sum_{k',q}V(k-k')V(k-q)G(k')G(q)
\left[\frac{dG_{\sigma}(q-k+k')}{dh}\right]_0 \nonumber \\
&&+\sum_{k',q}V(k-k')V(k-q)G(k')G(q-k+k')
\left[\frac{dG_{\sigma}(q)}{dh}\right]_0 \nonumber \\
\end{eqnarray}
and by re-arranging the 4-vectors indeces and using (\ref{gamma1})
and (\ref{gamma2}) the spin-vertex equation reduces to:
\begin{equation}
\label{gamma7}
\Gamma(k)=1+\sum_{k'}\left[\widetilde{V}(k,k')+I\right]G(k')^2\Gamma(k'),
\end{equation}
where the electron-phonon kernel $\tilde{V}(k,k')$ is:
\begin{eqnarray}
\label{gamma8}
\widetilde{V}(k,k')\!&=&\!V(k-k')\!\left[1\!+\!2\!
\sum_qV(k-q)G(q)G(q-k+k')\right] \nonumber \\
&+&\sum_qV(k-q)V(q-k')G(q)G(k+k'-q) .
\end{eqnarray}
A graphical representation of (\ref{gamma7}) and (\ref{gamma8})
in terms of Feyn\-man diagrams is shown in Fig. \ref{figvertexchi-nafl}.

In comparison
with Fig. \ref{figvertexchi-nca},
the kernel of the spin susceptibility resulting from
the vertex corrected self-energy is modified by additional electron-phonon
contributions represented by vertex and cross diagrams.
This set of diagrams is therefore quite similar to those encountered
in the theory of nonadiabatic superconductivity. They however differ
in the orientation of the electron propagators (particle-hole
rather than particle-particle contributions) leading to a different
momentum dependence of the cross term. The evaluation
of the kernel (\ref{gamma8}) can therefore be carried on by
following a scheme similar to the one already employed in
previous works.\cite{pietro1,sgp}

Let us first evaluate the zero-field self-energy entering the
electron propagators in Eq. (\ref{gamma8}). As discussed before,
the Coulomb self-energy provides just a constant term which
can be absorbed into the definition of the chemical potential.
The total self-energy reduces therefore to the electron-phonon one
which from Eq. (\ref{self4})
can be written in the compact form:
\begin{eqnarray}
\label{self5-2}
\Sigma({\bf k},n)=&-&T\sum_m\sum_{{\bf k}'}
|g({\bf k}-{\bf k}')|^2 D(n-m) \nonumber \\
&\times&\left[1+P({\bf k},{\bf k}';n,m)\right]
G({\bf k}',m) ,
\end{eqnarray}
where we have introduced the vertex function given by:
\begin{eqnarray}
\label{vertex1}
&&P({\bf k},{\bf k}';n,m)=
-T\sum_{l}\sum_{{\bf q}}|g({\bf k}-{\bf q})|^2 D(n-l) \nonumber \\
&&\times G({\bf q}-{\bf k}+{\bf k}',l-n+m)G({\bf q},l). 
\end{eqnarray}

In similar way as in non-crossing approximation,
we eliminate the angular dependence of the self-energy
by replacing the whole electron-phonon matrix
element, which takes into account the
vertex correction, by its average over
the Fermi surface:
\begin{eqnarray}
\label{ave3}
&&|g({\bf k}-{\bf k}')|^2\left[1+P({\bf k},{\bf k}';n,m)\right]
\rightarrow \nonumber \\
&&\langle \langle |g({\bf k}-{\bf k}')|^2\left[1+
P({\bf k},{\bf k}';n,m)\right]\rangle\rangle_{\rm FS} \nonumber \\
&&=g^2[1+\lambda P(Q_c;n,m)]
\end{eqnarray}
The analytic expression of $P(Q_c;n,m)$ and its explicit derivation
is provided in Appendix \ref{appa}.
The parameter $Q_c$ is a dimensionless momentum
describing the relevant momentum scale
of the electron-phonon scattering
process. In conventional metals, as the low temperature superconductors,
the maximum exchanged phonon momentum $q_c$ is about 
the Debye vector $q_{\rm D}$ and
$Q_c \sim q_{\rm D}/2k_F \sim 1$. It is well known, however,
that in systems with low density of charge carriers,
as fullerene's compounds and cuprates, the electrons scatter only with
charge fluctuations of long wavelength because of the strong electronic
correlation. In fact, in strongly correlated systems,
the electrons are surrounded by
giant correlation holes which suppress charge density fluctuations
with large momenta.\cite{grilli,zeyher,kulic}
The relevant exchanged phonon scale is thus
quite smaller than the Debye vector: $q_c < q_{\rm D}$ and
$Q_c = q_c/ 2k_F < 1$.
As a consequence the effective electron-phonon interaction
is suppressed for momenta larger than some cut-off $q_c$ which depends
on the filling factor and on the Coulomb repulsion.
We modelize this situation by introducing a sharp momentum cut-off
$q_c$. It can be used as a free parameter to distinguish
between weak and strong correlation cases, where
$q_c$ is small for strong correlated systems
and of order one for weakly interacting electrons.

As usual, the self-energy effects can be expressed in a convenient
form by means of the renormalized frequencies 
$W_n=\omega_n[1-\Sigma(n)/(i\omega_n)]$
obtained by integrating Eq. (\ref{self5-2}) over the energy.
In the vertex corrected theory
they fulfil the self-consistent relation:
\begin{eqnarray}
\label{self6}
W_n  =  \omega_n &-& \pi T_c\sum_m\lambda[1+\lambda
P(Q_c;n,m)]D(n-m) \nonumber \\
&\times&\frac{2}{\pi}
\arctan\left(\frac{E_F}{W_m}\right) .
\end{eqnarray}

By using the momentum independent self-energy (\ref{self6}),
the spin-vertex function (\ref{gamma7}) can be rewritten as:
\begin{equation}
\label{gamma9}
\Gamma ({\bf k},n)=1-T\sum_{{\bf k}',m}
\left[I+\widetilde{V}({\bf k},{\bf k}';n,m)\right] 
\frac{\Gamma ({\bf k}',m)}{[iW_m-\epsilon({\bf k}')]^2},
\end{equation}
where the nonadiabatic electron-phonon
spin kernel $\tilde{V}({\bf k},{\bf k}';n,m)$ 
is given by:
\begin{eqnarray}
\label{gamma10}
\widetilde{V}({\bf k},{\bf k}';n,m)&=&|g({\bf k}-{\bf k}')|^2D(n-m)\nonumber \\
&\times &\left[1+2 P({\bf k},{\bf k}';n,m)\right]
+C({\bf k},{\bf k}';n,m). \nonumber  \\
\end{eqnarray}
Here $P({\bf k},{\bf k}';n,m)$ is again the first vertex correction
given in Eq. (\ref{vertex1}) and $C({\bf k},{\bf k}';n,m)$ is the
cross correction whose the explicit expression is given below:
\begin{eqnarray}
\label{crossa}
&&C({\bf k},{\bf k}';n,m)=
T\sum_l\sum_{{\bf q}}|g({\bf k}-{\bf q})|^2
|g({\bf q}-{\bf k}')|^2\nonumber \\
&&\times \frac{D(n-l)D(l-m)}
{[iW_l-\epsilon({\bf q})]
[iW_{n+m-l}-\epsilon({\bf k}+{\bf k}'-{\bf q})]}.
\end{eqnarray}

Coherently with the approximations performed on the
self-energy and with the Ward's relation Eq. (\ref{ward}),
we evaluate $\Gamma ({\bf k},m)$ by replacing the kernel
(\ref{gamma10}) by its momentum average over the Fermi
surface:
\begin{equation}
\label{ave5}
\widetilde{V}({\bf k},{\bf k}';n,m) 
\rightarrow \langle \langle \widetilde{V}({\bf k},{\bf k}';n,m)
\rangle\rangle_{\rm FS} =\widetilde{V}(Q_c;n,m) ,
\end{equation}
where
\begin{eqnarray}
\label{ave6}
\widetilde{V}(Q_c;n,m)&=&\lambda D(n-m)\left[
1+2 \lambda P(Q_c;n,m)\right]  \nonumber \\
&+&\lambda^2 C(Q_c;n,m).
\end{eqnarray}
The explicit expression of $C(Q_c;n,m)$ can be also
found in Appendix \ref{appa}.

The final expression of the ladder vertex equation beyond
the adiabatic approximation is readily obtained from 
Eqs. (\ref{gamma9})-(\ref{ave6}). The result of the integration over the
electron energy gives:
\begin{equation}
\label{gamma10-2}
\Gamma (n)=1-\lambda T\sum_{m}\left[I+\widetilde{V}(Q_c;n,m)\right]
\frac{2E_F}{W_m^2+E_F^2}\Gamma (m).
\label{gammavertex}
\end{equation}
Finally, the spin-susceptibility in the vertex corrected
approximation is obtained by plugging Eq. (\ref{gamma10-2})
into Eq. (\ref{chi5}).

\section{Results}
\label{results}

We are now in the position to calculate the Pauli susceptibility $\chi$
and to evaluate the effects on $\chi$ of the electron-phonon interaction,
both in the non-crossing approximation and in the vertex corrected theory.
Of course, when the adiabatic parameter $\omega_0/E_{\rm F}$
or the electron-phonon coupling constant $\lambda$ are turned to zero
$\chi$ would reduce to the simple Stoner enhanced susceptibility:
\begin{equation}
\lim_{\lambda \rightarrow 0}\chi(T)=
\lim_{\omega_0/E_{\rm F}\rightarrow 0}\chi(T)
=\frac{\chi_0(T)}
{1-I N_0\frac{\displaystyle \chi_0(T)}
{\displaystyle \chi_{\rm P}}},
\label{stoner}
\end{equation}
where $\chi_0(T)$ is the free-electron Pauli susceptibility:
\begin{eqnarray}
\label{chi0}
\chi_0(T)&=&-2\mu_{\rm B}^2T\sum_n\sum_{{\bf k}}
\frac{1}{[i\omega_n-\epsilon({\bf k})]^2} \nonumber \\
&=&\chi_{\rm P}[1-2f(E_F)],
\end{eqnarray}
where $f(E_F)$ is the Fermi distribution function at $E_F$.

\begin{figure}
\centerline{\psfig{figure=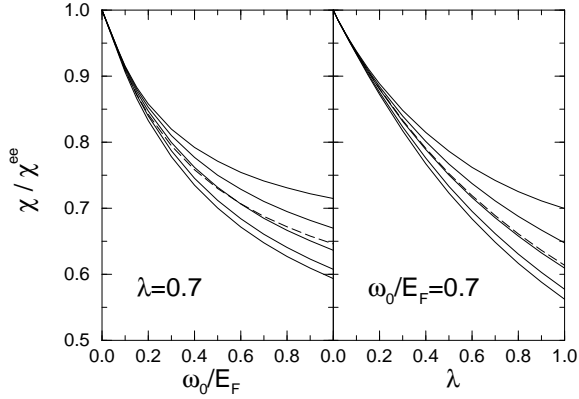,width=7.5cm}}
\caption{Spin susceptibility as function
of the adiabatic parameter $\omega_0/E_{\rm F}$
and of the electron-phonon coupling
$\lambda$  for $N_0 I=0.4$.
Dashed lines represent the
spin susceptibility for non-crossing approximation, solid lines
for the vertex corrected theory (from the lower to the upper line:
$Q_c=0.1$, $Q_c=0.3$, $Q_c=0.5$, $Q_c=0.7$, $Q_c=1.0$).}
\label{chi-vs-e-l}
\end{figure}

In the following we will denote the zero electron-phonon limit 
in Eq. (\ref{stoner}) as $\chi^{\rm ee}$ and it would be used as comparison
element to evaluate the effects of the electron-phonon interaction.

In Fig. \ref{chi-vs-e-l}  we plot
the total spin susceptibility (electron-electron + electron-phonon
scattering) as function of the electron-phonon coupling and of
the adiabatic parameter for zero temperature. Dashed lines are the results
obtained within the non-crossing approximation while the solid lines
are the data for the vertex corrected theory. For this latter case
we show the results for different values of the momentum cut-off
$Q_c$ ($Q_c=0.1$, $0.3$, $0.5$, $0.7$, $1.0$).

The first main result of Fig. \ref{chi-vs-e-l} is
that the inclusion of the electron-phonon coupling,
{\em in the nonadiabatic regime} $\omega_0/E_{\rm F} \geq 0$ yields
a sensible reduction of $\chi$ with respect to the
pure electronic spin susceptibility. 
As expected this effect vanishes as $\lambda \rightarrow 0$ 
(right panel) or
$\omega_0/E_{\rm F} \rightarrow 0$ (left panel). 
Note that both the non-crossing and vertex corrected theories
yield similar reduction. This is quite different from the situation
encountered in the superconducting pairing channel,\cite{cgps,pietro1,sgp}
where the effect of the vertex corrections is much stronger and highly
dependent on $Q_c$.

The results of Fig. \ref{chi-vs-e-l} suggest that
some care is needed in estimating the value of the bare density of states
from paramagnetic susceptibility measurements. Indeed, our analysis shows 
that $\chi$ is not simply related to the density of states by 
Eq. (\ref{stoner}). Namely, disregarding the electron-phonon effects
would lead to a substantial understimation of the bare density of states
from a spin susceptibility measurement as long as the electron-phonon 
interaction is in the nonadiabatic regime. 

\begin{figure}
\centerline{\psfig{figure=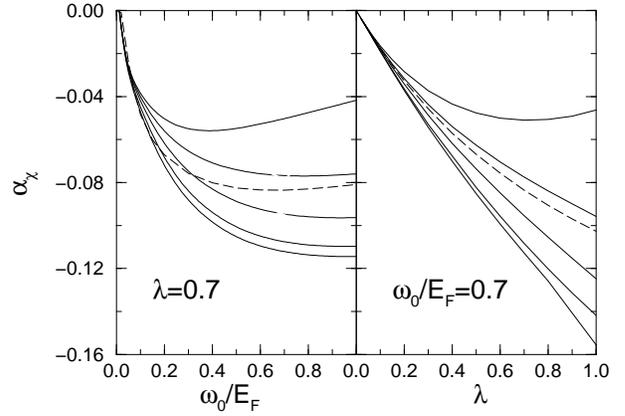,width=7.5cm}}
\caption{Isotope effect on the spin susceptibility as function
of $\omega_0/E_{\rm F}$ ($\lambda=0.7$, $N_0 I=0.4$) and as function of
$\lambda$ ($\omega_0/E_{\rm F}=0.7$, $N_0 I=0.4$).
Solid lines and dashed line 
as in previous captions.}
\label{figachi-el-nafl}
\end{figure}

A more clear signature of the nonadiabatic effects is
provided by the isotope dependence of the spin susceptibility.
In Fig. \ref{figachi-el-nafl} we report the numerical calculations
of the isotope coefficient $\alpha_\chi=-d\log\chi/d\log M$,
where $M$ is the ion mass, as a function of the adiabatic ratio
$\omega_0/E_F$ and of $\lambda$. Both the non-crossing (dashed lines)
and the vertex corrected (solid lines)
theories predict negative values of $\alpha_\chi$. Compared to the 
non-crossing data, the vertex corrected results show for small
values of $Q_c$ a stronger dependence on $\omega_0/E_F$ and $\lambda$
leading to $\alpha_\chi$ of about $-0.1$. The observation
of the isotope effect, which should be absent in metals 
fulfilling the ME framework, could be therefore a stringent evidence
of a nonadiabatic electron-phonon coupling. Note that a previous analysis of
experimental data of some superconducting properties has demonstrated 
the failure of the ME theory for Rb$_3$C$_{60}$, pointing out the
breakdown of Migdal's theorem in fullerides.
In this respect, a measurement of $\alpha_\chi$ in C$_{60}$ compounds 
would represent a direct and independent test of the relevance of
nonadiabatic electron-phonon interaction.

\begin{figure}
\centerline{\psfig{figure=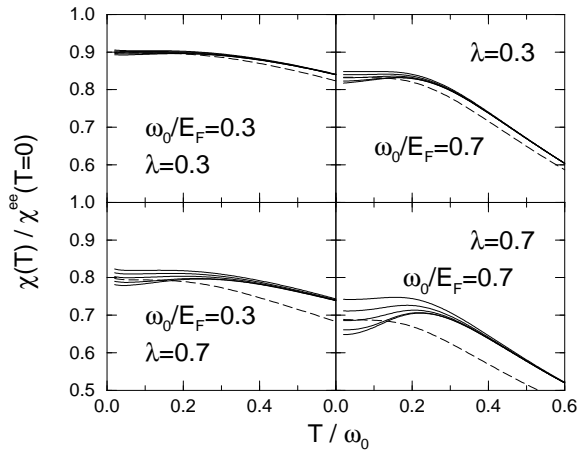,width=7.5cm}}
\caption{Temperature dependence of the spin susceptibility for
different values of $\omega_0/E_{\rm F}$ and
$\lambda$ and $N_0 I=0.4$. Solid lines and dashed line 
as in previous captions.}
\label{figchi-t-nafl}
\end{figure}

From a qualitative point of view,
the zero temperature behaviors of both the non-crossing and
vertex corrected theories give substantially similar results.
A interesting differentiation among the two approximations
arises in the temperature dependence of $\chi$. 
In Fig. \ref{figchi-t-nafl} we report the calculated temperature
dependence of the Pauli susceptibility for different values of 
$\omega_0/E_F$ and $\lambda$. The non-crossing approximation (dashed lines)
yields a monotone decreasing of $\chi$ as the temperature $T$
increases. This is basically due to the phonon cut-off in the ladder
equation for the susceptibility, Eq. (\ref{gammanca}). 
On the other hand, the vertex corrected $\chi$ (solid lines) has a 
richer temperature dependence which becomes more important as
$\lambda\omega_0/E_F$ increases. Starting from $T=0$, the basic feature is 
represented by an initial increase of $\chi$ with $T$
followed by a decreasing for larger temperatures. Although the decreasing
part is rather $Q_c$ independent, the initial increase of $\chi$ is
steeper for lower values of the momentum cutoff. Note for example
that for $\lambda=0.7$, $\omega_0/E_F=0.7$ and $Q_c=0.1$, 
at $T/\omega_0\simeq 0.2$ the susceptibility is enhanced by a 
8-9 \% of its value at $T=0$. Although this increase is rather small, it
is nevertheless an interesting feature since it is not related to
any increase of the lattice constant due to thermal expansion (the
calculations reported here in fact are done for constant volumes).
Experimentally in fact, $\chi$ at $T\simeq 300$ K is found to be larger
than its value at $T\sim 25$ K of $\sim 30$ \%
and $\sim 40$ \% for K$_3$C$_{60}$ and Rb$_3$C$_{60}$, 
respectively.\cite{tanigaki,robert}
Moreover recent data suggest a following decrease of $\chi$ in
K$_3$C$_{60}$ by increasing temperature.\cite{forro}
The initial increase of $\chi$ with temperature
is currently explained by a temperature enhancement of the
density of states at the Fermi level due to the thermal expansion
of the unit cell. Note however that a power law dependence of the
density of state on the lattice constant is not sufficient to
reproduce the observed increase of $\chi$, while a stronger dependence
like an exponential law needs a quite small value of the Stoner enhancement,
$(1-N_0 I)^{-1}\simeq 1.3$, to fit the experimental data of $\chi$.\cite{robert}
Such small Stoner enhancement is in contrast to recent Monte Carlo
calculations which estimate $(1-N_0 I)^{-1}\simeq 3$.\cite{arya} 
Although a detailed study of the effect of the thermal expansion on
$\chi$ is beyond the scope of this paper, the results of 
Fig. \ref{figchi-t-nafl} suggest that the electron-phonon contribution
could be an additional source for the temperature dependence of $\chi$.
Of course, a firmer evidence of the role of nonadiabaticity would
be the measurement of the temperature dependence of $\chi$ 
for a constant sample volume, in the spirit therefore of the constant volume
resistivity experiment of some years ago.\cite{vareka}

An alternative explanation of the non monotone temperature dependence
of $\chi$ has recently been attributed to possible effects
of mobile ions K$^+$.\cite{forro} 
This argument would predict however a finite isotope
effect on $\chi$ by alkali isotopic substitution and no carbon isotope
effect whereas in nonadiabatic theory an opposite trend is expected.

\section{Conclusions}
\label{concl}

In this paper the nonadiabatic theory
of the Pauli spin susceptibility in narrow-band systems
has been formulated in order to identify possible signatures
of nonadiabatic electron-phonon coupling in fullerides.
We have identified peculiar features that can be considered as hallmarks
of a relevant electron-phonon coupling {\em in nonadiabatic regime}.
In particular, an effective reduction of the spin susceptibility
by the nonadiabatic electron-phonon coupling has been found
in constrast to the conventional ME framework where no
electron-phonon renormalization is expected.
In addition, we predict a finite negative isotope effect on $\chi$
which we suggest as possible experimental test.
The Pauli spin susceptibility of $\chi$ also acquires in
nonadiabatic regime an anomalous temperature dependence in qualitative
agreement with the experimental data.

The investigation of nonadiabatic effects in C$_{60}$ compounds attracts
new and refreshed interest in the light of the recent experimental
indications of an unconventional phonon-based origin of the
superconducting pairing.\cite{cgps} 
The failure of the conventional theory of superconductivity
should not be surprising by considering that
one of the basic assumptions of Migdal-Eliashberg theory,
namely the adiabatic principle on which Migdal's theorem relies,
breaks down in fullerene compounds where phonon frequencies are
of the same order of the Fermi energy.

Aimed by these motivations, in the past years
we have developed the nonadiabatic theory of superconductivity
in narrow band systems that we propose as the theoretical framework to
properly describe fullerides and probably cuprates.
The inclusion of the nonadiabatic vertex corrections
arising from the breakdown of Migdal's theorem has permitted
to explain in a natural way some of the experimental features
of C$_{60}$ compounds
which appeared anomalous within the ME context.
For example, the low density of charge carriers can be regarded as
a characteristic element
since it is a direct by-product of the band narrowness
and as a consequence of the nonadiabatic regime. In addition,
the strong electron-electron interaction was shown to yield more
attractive pairing channels
than repulsive ones in {\em nonadiabatic} regime, and to play therefore
a positive role with respect to the superconducting onset.
More technical anomalies, as the reduction of $T_c$ in fullerides
by induced disorder, have also received a natural explanation
in the context of nonadiabatic theory.

The results of the present work
shed new light also on the anomalous $T_c$ {\rm vs.} $\chi$
dependence in ammoniated alkali doped fullerides.
In (NH$_3$)$_x$$A_3$C$_{60}$ indeed a sharp reduction of $T_c$,
together with a weak expansion of the lattice constant $a$,
is observed upon increasing of $x$.\cite{shimoda,maniwa,ricco}
In addition measurements of Pauli spin susceptibility show
that such a reduction of $T_c$ corresponds to an enhancement
of $\chi$.\cite{maniwa,ricco}
This experimental situation looks quite puzzling from the conventional
point of view. In ME theory indeed
an increase of the lattice constant $a$ is reflected in an enhancement of the
bare density of states at the Fermi level $N_0$.
Both the Pauli susceptibility $\chi \propto N_0$ and the critical temperature
$\log T_c \propto - 1/ N_0$ are therefore correspondingly expected to
increase. Taking into account Stoner enhancement would make even stronger
this trend. A theoretical explanation of this situation is
still a open issue.

\begin{figure}
\centerline{\psfig{figure=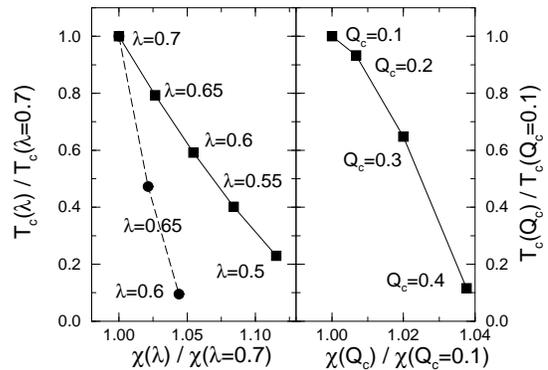,width=7.5cm}}
\caption{Critical temperature versus Pauli spin susceptibility
in nonadiabatic regime by varying $\lambda$ (left panel)
and $Q_c$ (right panel).}
\label{figchi-t-chi-lqc}
\end{figure}

The analysis of the Pauli susceptibility
here presented suggests however a new possible interpretation.
A crucial point is the observation that the Pauli susceptibility
in the nonadiabatic theory, apart from the Stoner factor, does not
give a direct probe of the electron density of states but it has
to be considered
as a quantity renormalized by electron-phonon effects.
Even more important is the fact that the renormalization itself depends
on electron-phonon properties, so that an increase or decrease of $\chi$
can be tune by electron-phonon properties. Namely, in the model considered
in this paper, the Pauli spin susceptibility $\chi$ acquires
a significant dependence on $\lambda$ and $Q_c$. Figure
\ref{chi-vs-e-l} shows an enhancement of $\chi$ by decreasing
$\lambda$ or $Q_c$ (in vertex corrected theory). From an intuitive point
of view such an enhancement of $\chi$ should be accompanied by a depletion
of $T_c$. We quantify this concept by calculating the superconducting
critical temperatures in non-crossing approximation and in vertex
corrected theory. 
We consider here $\lambda=0.7$, 
$Q_c=0.1$, $\omega_0/E_F=0.7$ and $N_0 I=0.4$,
and we study the $T_c$ {\rm vs.} $\chi$ dependence as function
of $\lambda$ (left panel) and $Q_c$ (right panel).
Numerical calculation are shown in 
in Fig. \ref{figchi-t-chi-lqc} for both the theories:
solid lines represent the vertex corrected one and the dashed line
the non-crossing approximation which does not depend on $Q_c$.
The results shown in Fig. \ref{figchi-t-chi-lqc} suggest
therefore that the driving element of the anomalous
$T_c$ {\rm vs.} $\chi$ behaviour could be the electron-phonon coupling
in nonadiabatic regime
and the weak lattice expansion of secondary importance.
A microscopic relation between ammonia intercalation
and possible modifications of the electron-phonon interaction is still
missing. The high critical temperature of 
(NH$_3$)$_4$Na$_2$CsC$_{60}$ however seems indicate that an important role
can be played by the displacement of the alkali cations from
the center of the interstitial sites and by the splitting
of the crystal field.

\begin{appendix}

\section{}
\label{appa}

Aim of this appendix is to provide the analytical evaluation
of the electron-phonon vertex correction $P({\bf k},{\bf k}';n,m)$
and cross function $C({\bf k},{\bf k}';n,m)$ and of their average over 
the momentum transfer.
The following derivation is an improvement of the previous
calculations presented in the appendix of Ref. \onlinecite{pietro1}
by explicitely including the self-consistent self-energy renormalization.
In order to do that a different procedure is employed.
Let us consider first the vertex correction, Eq. (\ref{vertex1}), 
reported below for convenience:

\begin{eqnarray}
&&P({\bf k},{\bf k}';n,m)=
-T\sum_l\sum_{\bf q}|g({\bf k}-{\bf q})|^2 D(n-l)
\nonumber \\
&&\times\frac{1}{iW_l-\epsilon({\bf q})}
\frac{1}{iW_{l-n+m}-\epsilon({\bf q}-{\bf k}+{\bf k}')},  
\label{p1}
\end{eqnarray}
We consider a model for the
electron-phonon matrix element which simulates the effects
of the strong electronic correlation.
Strong electron correlation has been shown in literature to
favour the forward electron-phonon scattering at
small transferred momenta.\cite{grilli,zeyher,kulic}
Following Ref. \onlinecite{pietro1},
we therefore schematize the electron-phonon matrix $|g({\bf q})|^2$ as:
\begin{equation}
\label{gmodel}
|g({\bf q})|^2=\frac{g^2}{Q_c^2}\Theta(Q_c-Q),
\end{equation}
where $Q=q/2k_F$, $Q_c=q_c/2k_F$ and $k_F$ is the Fermi wave vector.
$\Theta$ is the Heaviside step function and
$q_c$ is a momentum cut-off which depends on the degree of correlation in
the system: $q_c$ is smaller for more correlated materials.

The explicit inclusion of the self-energy in $W_n$
does not allow for an analytical
integration on the Matsubara frequencies.
In contrast with Ref. \onlinecite{pietro1}
we use thus the alternative procedure to evaluate first the integration
over the momenta leaving the summation over the frequencies untouched.
In evaluating the integration over the Brillouin zone, we employ
some basilar approximations which are justified by the dimensionless
cut-off $Q_c$ of Eq. (\ref{gmodel}) supposed to be small 
(strong correlation case).
In this context, we expand the electronic dispersion term 
$\epsilon({\bf p}-{\bf k}+{\bf k}')$ as:
\begin{equation}
\epsilon({\bf p}-{\bf k}+{\bf k}') \simeq 
\epsilon({\bf p})+2E_FQ\alpha \cos\phi,
\label{disp}
\end{equation}
where $E_F=v_F k_F$ and $\alpha$ and $\phi$ are the polar 
angles between ${\bf q}$ and
${\bf k}$. In writing Eq. (\ref{disp}),  we have assumed a isotropic system and set
$|{\bf k}|=|{\bf k}'|=|{\bf q}|=k_F$, $Q=|{\bf k}-{\bf k}'|/2K_F$.
The smallness of the second term of Eq. (\ref{disp}) is
enforced by the $\Theta$-functions of $g({\bf q})^2$ present
in Eq. (\ref{p1}) and in front of the vertex
function [see Eq. (\ref{ave3})].

In similar way, the leading order of the $\Theta$-function appearing
in Eq. (\ref{p1}) reads:
\begin{equation}
\Theta(q_c-|{\bf k}-{\bf q}|)\simeq \Theta(2Q_c-\alpha).
\label{tetta}
\end{equation}
Finally, the sum over the Brillouin zone can be expressed in terms of the same
angular coordinates $\alpha$, $\phi$:
\begin{equation}
\sum_{\bf p}
=\int_{-\pi}^{\pi}\frac{d\phi}{2\pi}
\int_0^\pi \frac{d\alpha \sin\alpha}{2}
\int_{-E_F}^{E_F}N(\epsilon) d\epsilon.
\label{angl}
\end{equation} 
By making use of Eqs. (\ref{disp}), (\ref{tetta}), (\ref{angl})
in Eq. (\ref{p1}) we obtain the expression for the electron-phonon
vertex function:
\begin{eqnarray}
&&P({\bf k},{\bf k}';n,m)=\frac{\lambda}{Q_c^2}
T\sum_l D(n-l)
\int_{-\pi}^{\pi}\frac{d\phi}{2\pi}
\int_0^{2Q_c} \frac{d\alpha \alpha}{2}\nonumber\\
&&\times\int_{-E_F}^{E_F} d\epsilon
\frac{1}{iW_l-\epsilon}
\frac{1}{iW_{l-n+m}-\epsilon-2E_F Q\alpha \cos\phi}, 
\label{p2}
\end{eqnarray}
where, coherently with the previous model, we have assume
a constant DOS $N(\epsilon)=N_0$.

It is convenient rewrite Eq. (\ref{p2}) in the
form:
\begin{eqnarray}
&&P({\bf k},{\bf k}';n,m)=\frac{\lambda}{Q_c^2}
T\sum_l D(n-l)
\int_{-\pi}^{\pi}\frac{d\phi}{2\pi}\nonumber\\
&&\int_0^{2Q_c} \frac{d\alpha \alpha}{2}
\frac{2E_F Q\alpha \cos\phi-i\left[W_l-W_{l-n+m}\right]}
{\left[2E_F Q\alpha \cos\phi\right]^2
+\left[W_l-W_{l-n+m}\right]^2}\nonumber\\
&&\int_{-E_F}^{E_F} d\epsilon
\left[\frac{\epsilon+iW_l}{\epsilon^2+W_l^2}
-\frac{\epsilon+2E_F Q\alpha \cos\phi+iW_{l-n+m}}
{\left(\epsilon+2E_F Q\alpha \cos\phi\right)^2+W_{l-n+m}^2}
\right].
\label{p3}
\end{eqnarray}

A very important role on the structure of the electron-phonon
vertex function is played by the term in the second line of
Eq. (\ref{p3}). It contains  a factor
$1/\left[(2E_F Q\alpha \cos\phi)^2
+(W_l-W_{l-n+m})^2\right]$ which is evidently nonanalitic
for $Q \rightarrow 0$, $W_l\rightarrow W_{l-n+m}$.
The different behaviour of this term with respect to the
opposite regimes $Q/\left[W_l-W_{l-n+m}\right] \ll 1$
or $Q/\left[W_l-W_{l-n+m}\right] \gg 1$ make arise the difference between
the static and dynamic limits and characterizes the main feature
of the momentum-frequency structure of $P$.
In contrast to that, the integrand in the third line of Eq. (\ref{p3})
is basically regular in the whole momentum-frequency space. 
Hence, in the small $Q$ expansion justified by the correlation cut-off
parameter $Q_c$, this latter term can be safely expanded at the
second order in $Q$ without losing important features of the vertex function.

By integrating on $\epsilon$ and expanding in $Q$
we are then left with:

\begin{eqnarray}
&&P({\bf k},{\bf k}';n,m)=-\frac{\lambda}{Q_c^2}
T\sum_l D(n-l)
\int_{-\pi}^{\pi}\frac{d\phi}{2\pi}\nonumber\\
&&\int_0^{2Q_c} \frac{d\alpha \alpha}{2}
\frac{A(n,m,l)-B(n,m,l)(2E_F Q\alpha \cos\phi)^2}
{\left[2E_F Q\alpha \cos\phi\right]^2
+\left[W_l-W_{l-n+m}\right]^2},
\end{eqnarray}
where 
\begin{eqnarray}
\label{afun}
A(n,m,l)&=&(W_l-W_{l-n+m})\left[\arctan\!
\left(\frac{E_F}{W_l}\right)\right.\nonumber \\
&&\left.-\arctan\!\left(\frac{E_F}{W_{l-n+m}}\right)\right], \\
\label{bfun}
B(n,m,l)&=&(W_l-W_{l-n+m})
\frac{E_FW_{l-n+m}}{[E_F^2+W_{l-n+m}^2]^2} \nonumber \\
&&-\frac{E_F}{E_F^2+W_{l-n+m}^2}.
\end{eqnarray}
The last two integrals on $\alpha$ and $\phi$ can now be performed
analytically giving:
\begin{eqnarray}
&&P({\bf k},{\bf k}' ;n,m)=\lambda P(Q,Q_c;n,m) \nonumber\\
&&= -\lambda T\sum_l D(n-l)
\left\{2B(n,m,l)\right.\nonumber\\
&&+\frac{A(n,m,l)
-B(n,m,l)\left[W_l-W_{l-n+m}\right]^2}
{(2E_FQQ_c)^2}\nonumber\\
&&\times\left.\left[
\sqrt{1+\left(\frac{2E_FQQ_c}{W_l-W_{l-n+m}}\right)^2}-1
\right]\right\}.
\end{eqnarray}
The average over the momentum $Q$ is finally obtained from Eq. (\ref{ave3}).
Also in this case, the integration can be done analytically and the
resulting averaged vertex function $P(Q_c;n,m)$ becomes:
\begin{eqnarray}
&&P(Q_c;n,m) = T\sum_l D(n-l)\left\{B(n,m,l)\right.\nonumber\\
&&+\frac{A(n,m,l)
-B(n,m,l)\left[W_l-W_{l-n+m}\right]^2}
{(2E_FQ_c^2)^2}\nonumber\\
&&\times\left[
\sqrt{1+\left(\frac{4E_FQ_c^2}{W_l-W_{l-n+m}}\right)^2}-1\right. \nonumber \\
&&\left.\left.-\ln\left(\frac{1}{2}\sqrt{1+\left(\frac{4E_FQ_c^2}{W_l-W_{l-n+m}}\right)^2}
\right)\right]\right\}.
\end{eqnarray}

The calculation of the cross function, Eq. (\ref{crossa}), follows basically
the same lines reported above. The main difference with respect to the
vertex function lies in the momentum dependence, which we handle in the same
way as reported in Ref. \onlinecite{cgps}. Therefore, in  Eq. (\ref{crossa}) 
we set $g({\bf k}-{\bf q})^2g({\bf q}-{\bf k}')^2\simeq
g({\bf k}-{\bf q})^2g({\bf k}-{\bf k}')^2$, which is an approximation valid
for small values of $Q_c$. Concerning the electronic dispersion, we
expand $\epsilon({\bf k}+{\bf k}'-{\bf q})$ in the following way:
\begin{eqnarray}
\epsilon({\bf k}+{\bf k}'-{\bf q})\simeq &&\epsilon({\bf q})+
E_F(1-Q^2)\alpha^2 \nonumber \\
&&-2E_FQ\sqrt{1-Q^2}\alpha\cos\phi,
\end{eqnarray}
where the angles $\alpha$ and $\phi$ have the same meaning as before.
The integration over the energy and the angles can now be performed
and, up to the leading order in the momentum transfer, the cross
function becomes independent of $Q=|{\bf k}-{\bf k}'|/2k_F$. 
Hence, for small values of $Q_c$, the momentum average leaves 
the cross function unaltered and, from Eqs. (\ref{ave5},\ref{ave6}), we obtain
\begin{eqnarray}
&&C(Q_c;n,m)=T\sum_l D(n-l) D(l-m)\left\{2B(n,-m,l)\right. \nonumber \\
&&+\frac{A(n,-m,l)-B(n,-m,l)(W_l-W_{l-n-m})^2}
{2E_F^2Q_c^2|W_l-W_{l-n-m}|} \nonumber \\
&&\left.\times \arctan\left(\frac{4E_FQ_c^2}{|W_l-W_{l-n-m}|}\right)\right\}
\end{eqnarray}
where the functions $A$ and $B$ are reported in Eqs. (\ref{afun},\ref{bfun}).

\end{appendix}

\end{document}